# Note on tax enforcement and transfer pricing manipulation

**Alex A. T. Rathke**

*School of Economics, BA and Accounting, University of São Paulo, Brazil*

*E-mail: alex.rathke@usp.br*

**Abstract:** This note proposes the segregation of independent endogenous and exogenous components of tax penalty probability to introduce a formal demonstration that enforcement and tax penalties are negatively related with income shifting.

## I. Introduction

Recent studies on tax avoidance emphasize the effects of tax enforcement against income shifting in multinational enterprizes (MNEs). They argue that enforcement creates a tax cost if noncompliance exists, and countries present different levels of fiscal stringency that affects MNEs' tax planning[1]. The Organization for Economic Co-operation and Development (OECD) also highlights tax enforcement to restrain transfer pricing manipulation (OECD, 2013), explaining that governments' enforcement influences the effectiveness of tax rules. In particular, the general premise is that enforcement increases the probability of imposition of fines whether MNEs manipulate transfer prices, forcing MNEs' compliance. Although current literature recognizes the relevance of tax enforcement, it typically takes the stringency effect as an assumption, and few studies effectively specify the penalty cost function in the analysis[2]. This note uses the model of tax costs of transfer pricing manipulation to verify the premise of tax enforcement effect on income shifting. The proposed model segregates the penalty probability in two independent components: an endogenous and an exogenous component with respect to MNE. The model embodies a specification of the endogenous probability in accordance with standard conditions in literature, and applies this specification to demonstrate formally that enforcement and penalties have a negative relation with income shifting. It shows that the magnitude of tax differential shapes the optimal level of transfer pricing manipulation, but enforcing tax effectiveness reduces this magnitude.

---

[1] See Lohse and Riedel (2013) and Beuselink et al. (2014) for recent empirical findings.

**II. The model**

Consider a MNE with two divisions in different countries ($i = 1,2$) producing ($x_i$) and selling ($s_i$) domestically. Division 1 exports part of its output ($m$) to division 2, charging a transfer price $p$. Using one set of books, MNE's pretax profits $\pi_i = R_i(s_i) - C_i(x_i)$ are

$$\pi_1 = R_1(s_1) - C_1(s_1 + m) + pm \quad \text{and} \quad \pi_2 = R_2(s_2) - C_2(s_2 - m) - pm$$

with $x_1 = s_1 + m$ and $x_2 = s_2 - m$. For an income tax rate $t_i$ on each country, MNE's after tax profits are $\Pi = (1 - t_1)\pi_1 + (1 - t_2)\pi_2$. If $t_1 \neq t_2$, MNE has incentive to shift profits from high-tax to low-tax country via intrafirm trades (Yoshimine and Norrbin, 2007): if $t_2 > t_1$, profits increase if $p$ increases – the *HighTP* case; if $t_1 > t_2$, profits increase if $p$ decreases – the *LowTP* case. In order to discourage income shifting, countries impose limitations in setting transfer prices and levy a penalty if MNEs do not comply with these rules. The widely accepted criterion focuses in the OECD guidelines, which determine that transfer prices must correspond to the arm's length price shaped by free trades.

---

[2] e.g. Amerighi (2008) specifies the penalty function, but does not limit upward threshold as 1.

Assume the harmed country $i$ settles a penalty $\Psi_i$ whether the difference between $p$ and the arm's length price $w$ causes profits to be shifted away from it. The probability $\alpha$ of imposition of $\Psi_i$ depends on the extent of the difference $p - w$. Additionally, assume that imposition of $\Psi_i$ is related with country $i$'s tax enforcement level $\theta_i$, evaluated in a standard range $0 \leq \theta_i \leq 1$, positively related with governments' taxation effectiveness and not influenced by MNE's decisions. For a unitary penalty $\varphi_i > 0$, the expected tax cost of transfer pricing manipulation is

$$\Psi_i = \varphi_i m \alpha \theta_i + 0 m (1 - \alpha) \theta_i = \varphi_i m \alpha \theta_i \quad (1)$$

MNE's objective function becomes $\phi = \Pi - \Psi_i$ ($i = 2$ for *HighTP*, and $i = 1$ for *LowTP*). For $\theta_i > 0$, transfer pricing manipulation implies $\Psi_i > 0$[3]. Thus, MNE maximizes profits when achieves equilibrium between reduced overall taxation and expected tax penalty.

### III. Specification of $\alpha$ and optimal transfer pricing manipulation

To design a fair specification for $\theta_i$ is quite a complex task because the assessment of tax enforcement involve country-level institutional factors that are difficult to measure[4]. Nonetheless, a specification for $\alpha$ appears to be simpler and is useful for the analysis. Standard model assumes the expected tax penalty as a continuous function $f(p - w)$,

---

[3] Transfer pricing manipulation implies $\phi_\Psi < 0$.
[4] Some studies create measures of tax enforcement based on rules' characteristics, e.g. Lohse *et al.* (2012).

twice differentiable and satisfying $f(p = w) = 0$, sign $f_p$ = sign$(p - w)$ and $f_{pp} > 0$[5].

Following Kant (1990), assume $P$ to be a limiting price that triggers penalty with certainty, i.e. $\alpha$ increases as $p$ gets closer to $P$. Henceforth, $\alpha$ can be specified as:

$$\alpha = \frac{(p-w)^r}{(P-w)^r} \quad (2)$$

where $r$ represents curve's convexity. Logically, $\alpha$ infers:

- for *HighTP*, $w < p < P$ implies $0 < \alpha < 1$, $w \geq p$ implies $\alpha = 0$, and $p \geq P$ implies $\alpha = 1$;

- for *LowTP*, $w > p > P$ implies $0 < \alpha < 1$, $w \leq p$ implies $\alpha = 0$, and $p \leq P$ implies $\alpha = 1$.

Equation (2) has interesting characteristics. First, it is continuous and satisfies the standard conditions for all $r > 1$ (Proof in Appendix). Second, it implies that $w$ is shaped by free trade forces. Tax authorities use $w$ as a parameter, but do not have power to influence it. Third, $P - w$ is the range of acceptable arm's length prices, so $p$ can be hold as tax compliant within it (OECD, 2013).

Specification of $\alpha$ as in Equation (2) allows the segregation of penalty probability in two independent components: $\alpha$ represents the endogenous component with respect to MNE, while $\theta_i$ represents the exogenous component. MNE can influence $\Psi_i$ because variations in $p$ cause changes in $\alpha$, despite of straight impacts on income shifting. On the other hand, tax authorities are able to increase penalty probability by enforcing audit procedures and strengthening their interpretation of what is an acceptable arm's length price, although these actions cannot influence $\alpha$ because it cannot affect directly the comparative parameters $w$ or $P$.

---

[5] (Double) subscripts denote first (second) derivatives with respect to indicated variables.

The equilibrium between the level of transfer pricing manipulation and the corresponding manipulation cost can be found differentiating $\phi$ with respect to $p$. Assuming $r = 2$ for simplification, the first-order condition is

$$\phi_p = (t_2 - t_1)m - 2\frac{(p-w)}{(P-w)^2}\varphi_i\theta_i m = 0 \quad (3)$$

Clearly, the difference between $p$ and $w$ denotes the transfer price manipulation. Rearranging Equation (3) leads to

$$(p-w) = \frac{(t_2 - t_1)}{2\varphi_i\theta_i}(P-w)^2 \quad (4)$$

The term $(t_2 - t_1)/2\varphi_i\theta_i$ reflects the influence of enforcement and unitary penalty on the tax differential, which is the income shifting motivation. Enforcement and penalties are negatively related with income shifting incentive and discourage transfer pricing manipulation, i.e. increase (decrease) in $\theta_i$ or $\varphi_i$ causes a decrease (increase) in $p - w$. The same relation is obtained for the more complex cases where both countries impose income taxation under residence-based principle and for partially owned affiliates[6]. From the analysis, the following proposition arises:

**Proposition:** *Assuming tax costs of transfer pricing manipulation as $\Psi_i = f(\theta_i, \varphi_i)$, tax enforcement $\theta_i$ and tax penalty $\varphi_i$ are negatively related with income shifting.*

**Proof:** See Appendix.

---

[6] Models of residence-based taxation and partially owned affiliates show the same negative relation between income shifting and tax enforcement (and penalty). Unpublished analysis is available from the author on request.

Equation (4) implies that the extent of tax differential determines the amount of MNE's price manipulation, but countries are able to toughen tax compliance actions in order to discourage this strategy. Higher enforcement levels and greater tax fines generate two effects. First, increasing $\theta_i$ causes probability of imposition of unitary penalty $\varphi_i$ to increase; as a consequence, the optimal $p$ get closer to $w$, and MNE is forced to reduce transfer pricing manipulation (be more tax compliant) in order to restore equilibrium; the same occurs when $\varphi_i$ increases. This is called "spillover effect", being an increase in compliance independent of revenues generated directly from tax audits themselves (Alm, 2012). Second, Equation (4) suggests that countries could manage both $\varphi_i$ and $\theta_i$ with intentions to recover evaded tax revenues, i.e. more tax prosecutions and greater fines effectively reduce MNE net profits. This is called "deterrent effect" of higher enforcement.

Equation (4) also shows that the range of acceptable arm's length prices is positively related with transfer pricing manipulation: the extent of $(P - w)^2$ represents a whole spectrum of alternatives where $p$ can be approved for tax purposes[7], and MNE may hold defense chances against tax inspections whether $p$ does not reach $P$.

**IV. Conclusion**

This note introduces a formal demonstration that enforcement and tax penalties are negatively related with income shifting. The extent of tax differential defines the optimal size of transfer pricing manipulation, but strengthening tax effectiveness reduces this extent. While there are few studies that determine formally the function of penalty cost, this note proposes the segregation of independent endogenous and exogenous components for the analysis. The specification of the endogenous probability presented here is expected to be more comprehensible with respect to MNEs' income shifting decisions, since it aims to approximate the realistic link between MNEs' transfer prices and arm's length criterion. Yet, the model maintains function's premises as stated in literature and infers results in line with empirical findings. This analysis intends to provide a simple and credible enhancement to existing literature and to spark further insights in income shifting research.

---

[7] Expanding (compressing) range $(P - w)^2$ formally creates more (less) opportunities for transfer pricing manipulations. Nonetheless, because $w$ is the main parameter for valuation of $p$, changes in $w$ generate different results for *HighTP* and *LowTP*.

**Appendix**

**Proof of conditions satisfied by $\alpha$.** Tax literature agrees on the following assumptions: expected tax penalty $= f(p - w), f(p = w) = 0$, sign $f_p$ = sign$(p - w)$ and $f_{pp} > 0$ for all $p \neq w$. Equation (2) shows that, for all $r > 1$:

$$\alpha(p = w) = 0$$

$$\alpha_p = \frac{r(p-w)^{r-1}}{(P-w)^r} \quad \begin{cases} > 0 \text{ for } HighTP \\ < 0 \text{ for } LowTP \end{cases} \quad \text{and}$$

$$\alpha_{pp} = (r^2 - r)\frac{(p-w)^{r-2}}{(P-w)^r} > 0 \quad \text{for both } HighTP \text{ and } LowTP$$

**Proof of Proposition.** Since $\theta_i$ and $\varphi_i$ run for the same direction within $\Psi_i$, let $G_i$ represent the effect of both variables ($\Psi_i = m\alpha G_i$). Assume $\alpha$ as specified in Equation (2). For the equilibrium in Equation (4), taking $(p - w) = f(G_i)$, the first-order condition with respect to $G_i$ gives

$$(p-w)_G = -\frac{\frac{(t_2 - t_1)}{rG_i^2}(P-w)^r}{(r-1)\left[\frac{(t_2 - t_1)}{rG_i}(P-w)^r\right]^{1-\frac{1}{r-1}}} < 0$$

for both *HighTP* and *LowTP*. If $\alpha = 1$ ($p \geq P$ in *HighTP* or $p \leq P$ in *LowTP*), manipulation cost $\Psi_i$ is not a function of $p$. Differentiating $\phi$ with respect to $G_i$ gives $\phi_G = -\Psi_G < 0$.